\documentclass{aa}
\usepackage{graphicx}
\begin{document}

\newcommand{\kms}{km~s$^{-1}$}
\newcommand{\cm}{cm$^{-2}$}
\newcommand{\lya}{Lyman~$\alpha$}
\newcommand{\lyb}{Lyman~$\beta$}
\newcommand{\za}{$z_{\rm abs}$}
\newcommand{\ze}{$z_{\rm em}$}
\newcommand{\nhi}{$N$(H~I)}

\def\ltsima{$\; \buildrel < \over \sim \;$}
\def\simlt{\lower.5ex\hbox{\ltsima}}
\def\gtsima{$\; \buildrel > \over \sim \;$}
\def\simgt{\lower.5ex\hbox{\gtsima}}
\def\arcs{$''~$}
\def\arcm{$'~$}

   \title{Unusual Metal Abundances in a Pair of Damped Lyman Alpha Systems at 
$z \sim 2$.\thanks{The work presented here is based on data obtained 
with UVES at the VLT, program 267.A-5707. }
}


   \author{S. L. Ellison\inst{1}
          \and
          S. Lopez\inst{2}
          }

   \offprints{S. Ellison}

  \institute{European Southern Observatory, Casilla 19001, Santiago 19, 
        Chile\\
              \email{sellison@eso.org}
        \and
	Departamento de Astronom\'{\i}a, Universidad de Chile, Casilla 36-D,
          Santiago, Chile \\
	 \email{slopez@das.uchile.cl}
}


   \date{Received / Accepted}

\abstract{We present high resolution spectroscopic
observations of two neighbouring damped \lya\ systems (DLAs) 
along the same line of sight towards B2314$-$409.  Due to their
separation ($v \sim$ 2000 \kms) and the high spectral resolution of the data,
it is possible to fit not only the weak metal transitions, but also
the separate H~I absorption profiles.   This has permitted, for
the first time, a detailed study of metal abundances in two neighbouring
galaxy-scale absorbers.
The two DLAs have \za\ = 1.8573 and 1.8745 and have column 
densities $\log$ \nhi\ = 20.9$\pm 0.1$ and 20.1$\pm 0.2$ respectively.  
We have determined abundances for a range of chemical elements,
and find that \textit{both} absorbers towards
B2314$-$409 have low $\alpha$/Fe-peak abundances compared with other known
DLAs.  This indicates that not only has the 
recent star formation history of these absorbers been relatively
passive, but that the group environment, or some other external
factor, may have influenced this.
}

\maketitle
%

\section{Introduction}

Since the discovery that intermediate
redshift ($z \sim 0.3$) clusters exhibit a relative over-abundance of
blue member galaxies compared with the local population (Butcher \& 
Oemler 1978), extensive work has investigated the evolution of
the cluster environment and compared it with the field population
(e.g. Poggianti et al. 1999 and references therein).  From these 
recent spectroscopic studies of $z \sim 0.4$
clusters, it has been established that
star formation is generally suppressed in these rich
environments, but that post-starburst (E + A) galaxies make up $\sim$ 20\% of
the cluster population (Dressler et al. 1999).  In addition, there appears
to be a radial star formation rate gradient in clusters that is independent of 
the morphology-density relation, such that galaxies with
the most recent star formation episodes occur farther out (Balogh et al. 1999).
Once accreted into the cluster, active star formation seems to be swiftly
quenched (Dressler et al. 1999) and continues at a relatively low
rate (e.g. Couch et al. 2001).

\begin{figure*}
\centerline{\rotatebox{270}{\resizebox{6.0cm}{15.0cm}
{\includegraphics{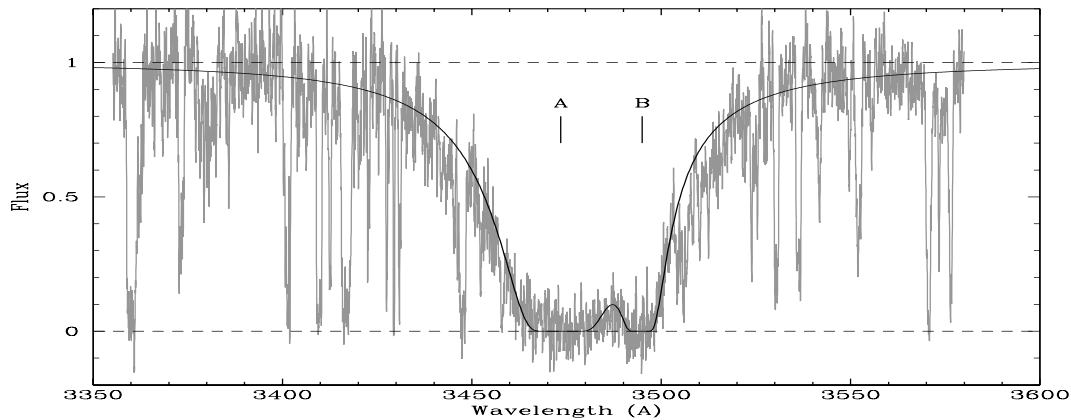} }}}
\caption{\label{HI_fit} Profile fit to the double DLA towards B2314$-$409.
The two absorbers have $\log$ \nhi\ = 20.9 at \za\ = 1.8573 (DLA A) and 
$\log$ \nhi\ = 20.1 at \za\ = 1.8745 (DLA B).}
\end{figure*}

However, these informative surveys have not been extended
beyond $z \sim 1$ due to the lack of good quality spectroscopic data at these
redshifts, although wide field surveys at X-ray, optical and near-IR
wavelengths have detected clusters out to $z \sim 1.3$ (e.g. Rosati et al.
1998).   At earlier epochs, the study of Lyman break galaxies (LBGs)
has permitted the discovery of large galaxy overdensities 
at $z \sim 3$ (Steidel et al. 1998).  
However, it is important to bear in mind that
being such biased tracers of matter, LBGs are very different from
typical cluster galaxies at low redshifts.  We are left, therefore, with a
significant gap in our knowledge of groups and clusters of galaxies
between $1 \simlt z \simlt 3$.  In particular, this leaves open many
issues involving the early evolution of galaxy groups.  For example,
at what stage does the environment start to affect the star formation
of the individual galaxies and is the activity boosted prior to being
truncated?  

One of the most promising
techniques for detecting representative galaxies at $z \simgt 1.5$ is
using QSO absorption lines, although the possibilities for studying
clusters of absorbing galaxies is more limited. 
Nonetheless, some observations of high column density absorbers,
in particular damped Lyman alpha
systems (DLAs), along multiple lines of sight have been supplemented
with Lyman break and narrow band \lya\ imaging to show
that DLAs can reside in galaxy concentrations out to $z \sim 3.5$
(e.g. Francis \& Hewett 1993; Francis, Woodgate and Danks 1997; 
Ellison et al. 2001), although there is currently no evidence
that DLAs cluster strongly with LBGs (Gawiser et al. 2001).  
In addition, the presence of metal line profiles with components
separated by many hundreds of \kms\ provides further evidence that DLAs may
have near neighbours (e.g. Pettini et al. 1999; Prochaska \& Wolfe 1999). 
However, due to the difficulty in determining the \nhi\ for individual
components, it has so far not been possible to perform full abundance analyses
of these proximate absorbers.

Here we present high resolution spectroscopic observations
of a DLA pair (i.e. two proximate absorbers in the same line of sight)
for which the abundances of the individual galaxies can be studied in detail
(Sect. 2) \footnote{We note that according to the historical
definition, only one of these absorbers is technically a DLA, despite
the presence of clear, broad damping wings.}.  In addition to determining
column densities for several metal line transitions, the UVES spectra
presented here have permitted us to resolve the two \lya\ lines, allowing us
to determine values for \nhi\ and therefore calculate abundances (Sect. 3).   
Various explanations for the unusual abundances exhibited by this DLA pair
are discussed (Sect. 4), including
dust and photo-ionization.  Since the chance alignment of two DLAs in 
single sightline is small, we also consider the possibility that
this double absorber is part of some galaxy structure at $z \sim 2$
and therefore whether their unusual abundances may be attributed to their
environment.

\section{Observations and Data Reduction}

Discovered as part of the CORALS survey for DLAs, the absorbers
towards B2314$-$409 were
identified by Ellison et al. (2002) as a potential DLA pair, based
on the asymmetric H~I profile and presence of metals separated by
$\sim$ 2000 \kms.   Four hours of Director's Discretionary time with UVES
were granted to observe B2314$-$409 at high spectral resolution in 
order to determine more accurately the DLA profile and metal abundances.
These observations were carried out on August 1 2001 with the dichroic
390+564 setting on UVES, the echelle spectrograph at the VLT, 
providing almost continuous wavelength coverage
from 3300 to 6650 \AA.  The data were reduced using the UVES pipeline,
a detailed description of which can be found in Ballester et al. (2000).
Once extracted, the individual frames ($R \sim 43\,000$) were
corrected to a vacuum heliocentric scale and combined with a weight
proportional to their S/N which varied between 10 and 25.
Finally, regions of the spectrum with absorption lines were normalised 
by dividing through by a spline function fitted through the QSO 
continuum.

\begin{figure}
\centerline{\rotatebox{0}{\resizebox{9.0cm}{!}
{\includegraphics{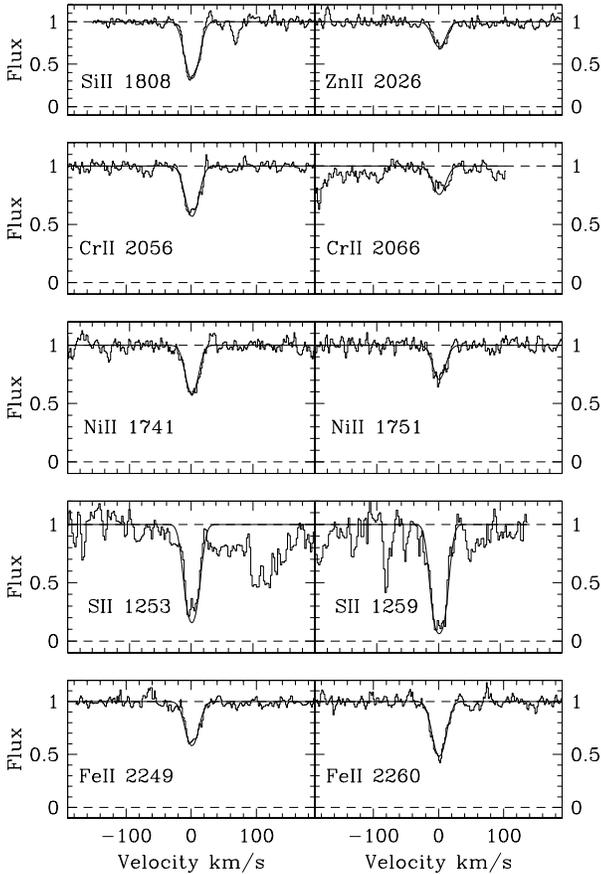} }}}
\caption{\label{metalsA} Unsaturated metal lines associated with DLA A with
Voigt profile fits overlaid. The velocity scale on each panel
is relative to \za\ = 1.8573.}
\end{figure}

\begin{figure}
\centerline{\rotatebox{0}{\resizebox{9.0cm}{!}
{\includegraphics{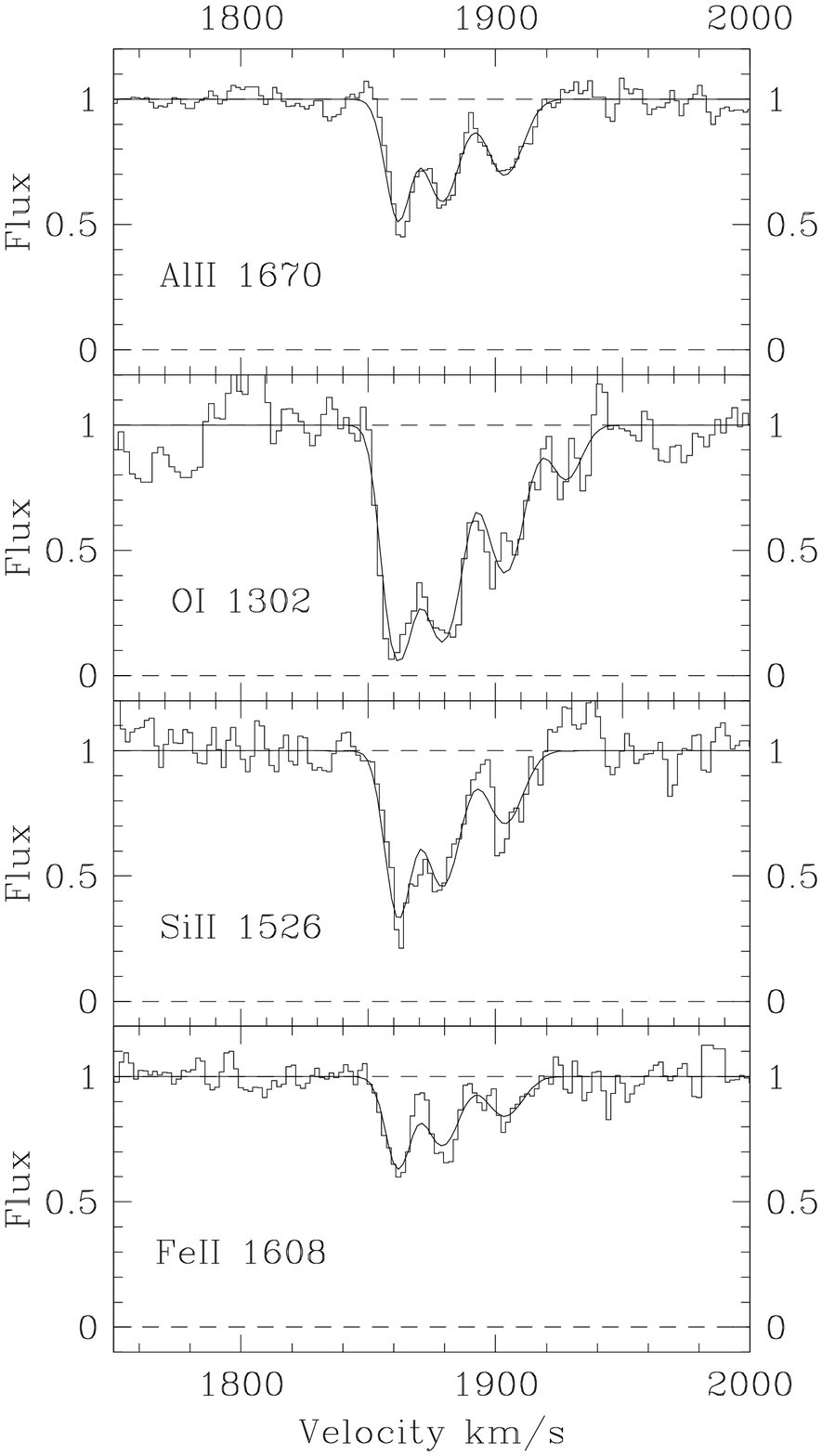} }}}
\caption{\label{metalsB} Unsaturated metal lines associated with DLA B with
Voigt profile fits overlaid. The velocity scale on each panel
is relative to \za\ = 1.8573.}
\end{figure}

The normalised section of the spectrum containing the DLA absorbers is
shown in Fig. \ref{HI_fit} together with a profile fit for the two
systems.  The redshifts and H~I column densities determined for the absorbers
are \za\ = 1.8573 and log \nhi\ = 20.9$\pm 0.1$ for DLA A and 
\za\ = 1.8745 and log \nhi\ = 20.1$\pm 0.2$ for DLA B, corresponding
to a proper separation of 11 Mpc $h^{-1}$ (H$_0$=70 \kms Mpc$^{-1}$,
$\Omega_M = 0.3$, $\Omega_{\Lambda} = 0.7$). 
These fits are constrained mostly by the shape of the damping wings,
although smoothing the data enhances the flux recovery between the
troughs that is already discernible in Fig. \ref{HI_fit}. A variety of metal
lines associated with these DLAs was identified and fitted with 
VPFIT\footnote{Available at http://www.ast.cam.ac.uk/\~{}rfc/vpfit.html }
to determine abundances.  As is common practice, we fixed the $b$-values
and redshifts of each absorption component, allowing only the column
density to vary.  Despite the fact that the same transitions 
were covered for both DLAs, the difference in their metallicities and H~I
column densities means that different transitions had to be used to determine
abundances in each case.  Typically, the transitions that were saturated in
DLA A provided good profile fits for DLA B, whilst the useful transitions
fitted in DLA A were not detected in DLA B.  Table 1 details the component
parameters determined for the both DLAs (shown in Figs \ref{metalsA} and
\ref{metalsB}) and Table 2 lists the total column densities and abundances.

\section{Abundances and Kinematics}
 
The abundances determined from Voigt profile fitting for both DLAs
are somewhat unusual, as we discuss in more detail below.  
An interesting possibility is that these
absorbers may be part of a large structure, for example
a proto-cluster or galaxy filament.  If so, the main
environmentally driven effects we may expect to witness will most likely
impact upon the gas kinematics and chemical abundances of the two
systems.  

\begin{table*}
\begin{center}
\caption{Voigt profile fit parameters for DLAs A and B towards Q2314$-$409. }
\begin{tabular}{ccccccccccc}\hline \hline
Cloud &Redshift & $b$& 
\multicolumn{8}{c}{Log$_{10}$ N(X)}\\ & &
 & Fe~II &  Zn~II & Cr~II & Al~II & Si~II & O~I & Ni~II & S~II\\ \hline
DLA A & & & & & & & & & & \\
1 & 1.857311  & 12.98  & 15.08 & 12.52 & 13.38 &... &15.41& ...&13.84&
15.10 \\ \hline
DLA B & & & & & & & & & & \\
1 &1.875032  & 5.02 &13.34 &...  &... &11.99 &13.43&14.41&...&...\\
2 &1.875197  & 7.74 &13.31 &...  &... &12.01 &13.38&14.28&...&...\\
3 &1.875431  & 8.42 &13.07 &...  &... &11.87 &13.05&13.95&...&...\\ 
4 &1.875661  & 6.82 & ... & ...  &...& ...&  ...&  13.32 &...&...\\\hline
\end{tabular}
\end{center}
\end{table*}
 
\begin{figure}
\centerline{\rotatebox{0}{\resizebox{10.0cm}{!}
{\includegraphics{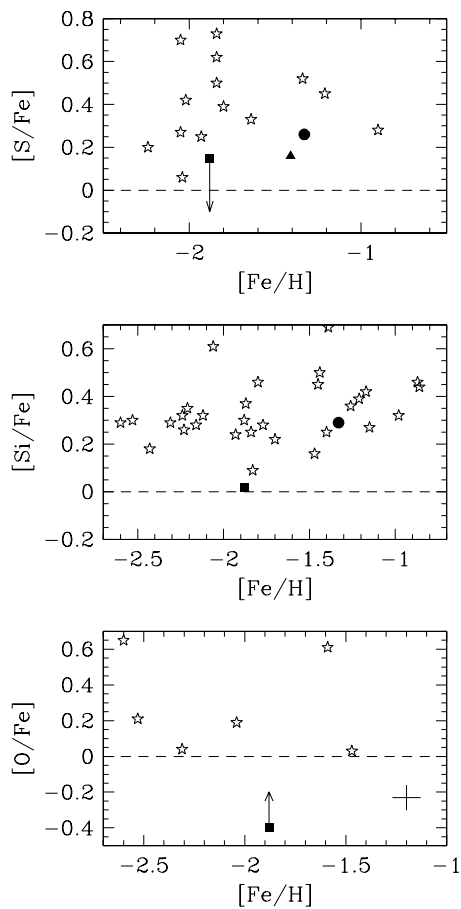} }}}
\caption{\label{alpha} Relative abundances of $\alpha$/Fe-peak elements.
In all panels, open stars are literature values (Molaro et al. 2001;
Outram et al. 1999; Lu et al. 1998; Ge, Bechtold \& Kulkarni 2001;
Pettini et al., in prep; Lu et al. 1996;
Centuri\'{o}n et al. 2000; Prochaska et al. 2001), the filled triangle is
for the DLA found towards Q0201+1220 in a galaxy group at $z \sim 3.4$ 
(Ellison et al. 2001), filled circles 
and squares represent values determined for DLAs A and B respectively.
The typical error bar is shown in the lower right corner of the bottom
panel.} 
\end{figure}

The overall metallicity of both DLAs is typical
of those measured at this redshift (Pettini et al. 1999), indicating
that there is no particular evidence of extended periods of
either highly enhanced or 
suppressed star formation over the span of each galaxy's
star-forming life.  However, we note that in documented DLAs
there is a large observed spread of metallicities at a given redshift,
so that to notice
a marked difference in [Zn/H] or [Fe/H] would require a very pronounced
effect.  Next we consider the abundances of $\alpha$ elements which, when
compared with Fe-peak elements, provide clues to the history of
star formation in the galaxy.  In DLA A, [S/Zn]
is roughly solar (see Table 2), Zn being the usual Fe-peak 
element of choice because of its non-refractory nature.  However, since
there are very few [S/Zn] measurements, in Fig. \ref{alpha} we plot
[S/Fe] in order to facilitate a useful comparison with literature values.  
Combined with the lower limit [S/Fe] $<0.15$ in DLA B, Fig.
\ref{alpha} reveals that both absorbers have relatively low S/Fe ratios
compared with other known DLAs.  Note that although the fit to Fe~II
$\lambda$1608 in DLA B appears somewhat poor (due to fixing the $b$-values),
allowing a completely free fit to the data results in only a 0.01 dex
change in $N$(Fe).
The abundance ratios plotted in Fig \ref{alpha} relative to Fe will
all require some dust correction, which will be different from
system to system (see next section).  
However, we nevertheless note that the [Si/Fe] ratio
for B is lower than any in the UCSD DLA database (Prochaska
et al. 2001) and the large compilation of Lu et al. (1996)\footnote{Since
several of the Fe abundances quoted by Lu et al. (1996) are based on
interpolation between upper and lower limits, we only plot the firm
detection in Fig. \ref{alpha}, taking into account the required 
corrections for updated $f$-values.}.  Although the [Si/Fe] ratio
in DLA A is fairly typical (c.f. its low [S/Fe]), this could be due,
at least in part, to some dust depletion as implied from [Zn/Cr] = +0.18.
Furthermore, for DLA B we determine a very low [O/Fe] = $-0.40$, 
which would be further reduced if there was some correction to be
made to Fe due to dust.  In reality, however, [O/Fe] = $-0.40$ is 
probably a $lower$ limit because of mild saturation of the O line
(see next section).  Overall, both 
DLAs exhibit relatively low $\alpha$/Fe-peak abundances,
although not excessively so, given the uncertainties.

Although DLA A appears to be a single component from unsaturated lines,
stronger transitions such as Si~II $\lambda$1526 and Al~II $\lambda$1670 
reveal this
system to have a somewhat more complicated multi-component structure.
In fact, both DLAs have absorption profiles that extend over approximately
100 \kms, a velocity not atypical compared with other damped systems
(Prochaska and Wolfe 2001).  It therefore appears that the interstellar
gas has not undergone significant disruption.  The few observations of 
other absorbers in high redshift galaxy groups provide mixed results
with regards to kinematics.
Q0201+1120 has a velocity spread of 300 \kms\ consisting
of many components (Ellison et al. 2001).  Similarly, the
DLA at \za\ = 2.38 towards B2138$-$4427 has components over 200 \kms\ 
(C. Ledoux, private communication), but the possible LLS in the same
group towards B2139$-$4434 has only a $\sim$ 60 \kms\ spread 
(V. D'Odorico, private communication).

\begin{table}
\begin{tabular}{ccccc}\hline \hline
  & \multicolumn{2}{c}{DLA A} & \multicolumn{2}{c}{DLA B} \\
X & N(X) & [X/H]& N(X) & [X/H] \\ \hline
Fe & 15.08$\pm0.10$ & $-1.33\pm0.14$ & 13.73$\pm0.13$ & $-1.88\pm0.24$\\
Si & 15.41$\pm0.10$ & $-1.04\pm0.14$ & 13.79$\pm0.10$ & $-1.86\pm0.22$\\
Cr & 13.38$\pm0.08$ & $-1.20\pm0.13$ & $<12.17$ & $<-1.61$\\
Zn & 12.52$\pm0.10$ & $-1.02\pm0.14$ & $<11.56$ & $<-1.19$\\
Ni & 13.84$\pm0.08$ & $-1.31\pm0.13$ & $<12.70$ & $<-1.65$\\
S  & 15.10$\pm0.15$ & $-1.07\pm0.18$ & $<13.64$ & $<-1.73$ \\ 
O  &  ... & ... & 14.75$\pm0.12$ & $-2.28\pm0.22^a$ \\ 
Al & ... & ... & 12.44$\pm0.08$ & $-2.14\pm0.22$ \\ \hline
\end{tabular}
\caption{\label{dla_tab}Abundance measurements (and 3$\sigma$ upper limits)
for DLAs A and B towards B2314$-$409. $^a$ -- Probably a lower limit due
to mild saturation of the O line.}
\end{table}

\section{Possible Explanations for Unusual Abundance Ratios}

The results of Fig. \ref{alpha} are striking.  It is particularly
noteworthy that \textit{both} DLA A and B exhibit unusually low
$\alpha$/Fe-peak ratios, as does the DLA in the group towards
Q0201+1120 (Ellison et al. 2001).  However, before drawing the conclusion
that this is evidence that the environment has a significant impact
on star formation histories at $z \sim 2$, we explore other factors
that may affect our abundance determinations.

We first consider the possibility of partial photo-ionization.
This is unlikely to be an issue for the relatively high \nhi\ DLA
A, but may have an effect on DLA B.  Evidence that low \nhi\ DLAs
and sub-DLAs may be increasingly affected by photo-ionization
comes from the increasing fraction of $N$(Al~III)/$N$(Al~II) found
by Vladilo et al. (2001).   However, the upper limit we determine for the
ratio of $N$(Al~III)/$N$(Al~II) $< -0.91$ is significantly $lower$ than
that predicted for a log \nhi\ = 20.1 from the trend found by 
Vladilo et al. (2001).  This provides direct
evidence that photo-ionization does not have a significant impact
on these abundance determinations.  Specifically, in the case of [O/Fe],
the calculations of Vladilo et al. (2001) show that Fe~II is converted 
to Fe~III as effectively as O~I is ionized to O~II, so that partial
ionization is highly unlikely to be the reason for the low O abundance. 
In fact, according to Vladilo et al. (2001), a ionization correction would
have a greater effect on [Si/Fe] pushing it to a $lower$ value and
enhancing the effect seen in Fig. \ref{alpha}. 

Next, we consider the effects of saturation.  As mentioned in the previous
section, it is plausible that the O~I $\lambda$1302 transition is mildly
saturated, requiring an upward revision of N(O).  In fact, it is possible
to achieve a fit with the same $\chi^2$ statistic if N(O) is increased
by 0.4 dex and the $b$-values allowed to vary freely.  A reduction of
the $b$-values by only $\sim$ 1.0 \kms\ in the two strongest components 
is sufficient to achieve this.  The additional corollary of increasing
$N$(O) is to bring the O/Si ratio in closer agreement with solar values.
On the other hand, the high spectral resolution
of these data make the possibility of `hidden' saturation in other
absorption lines highly unlikely.  Indeed, the excellent agreement
between column densities of Fe and Ni determined by using transitions 
with different oscillator strengths shows that there is no hidden
saturation effect for the other absorption lines used here.

Finally, we consider the possibility of atypical amounts of dust depletion.
The low [S/Fe] abundance shown in Fig. \ref{alpha} may be due
to anomalously low amounts of dust in these DLAs, although this is
not supported by the moderate [Zn/Cr] ratio in DLA A.  However,
dust is even more unlikely to be the explanation for low [Si/Fe]
ratios, since this would require that Si be more depleted than Fe,
contrary to what is observed locally (Savage \& Sembach 1996).  

\section{Discussion}
 
We have presented high resolution spectroscopic observations
of a pair of DLAs separated by $\sim 2000$ \kms\ ($\sim$ 11 Mpc$h_{70}^{-1}$)
at \za\ $\sim 2$ and propose the possibility that these absorbers
may form part of an extended structure.  
Absorbing structures of similar size at high $z$ have previously
been inferred from multiple lines of sight, for example the posited
super-cluster at $z \sim 2$ towards Tol 1037$-$2704 (Sargent \& Steidel 1987).
However, this is the
first time that the abundances of two proximate DLAs in a single line of sight 
have been studied in detail.  One other multiple
system, a triple DLA towards CTQ~247 (Lopez et al. 2000), has been 
recorded in the literature and a program has been initiated to
study its abundances.

Although we have found that the metallicity of
both DLAs is typical of that measured in other such systems at the
same redshift, there is evidence from their unusual abundance ratios
that these absorbers may have experienced somewhat different
star formation histories than most other documented DLAs.  
This evidence comes from the generally
low $\alpha$/Fe-peak abundances of \textit{both} DLAs, and the DLA
in the group towards Q0201+1120 (Ellison et al. 2001), suggestive that 
their global environment may have played some role in their evolution.  
Despite the possible effects of saturation (in the case of O~I $\lambda$1302), 
the low [Si/Fe] in DLA B and [S/Fe] in DLA A
compared to the large number of literature values remains convincing
evidence that star formation history may be responsible for these ratios.  
Mild dust depletion may explain the moderate [Si/Fe] in DLA A,
in which case [S/Fe] would be further decreased.
Such low $\alpha$/Fe-peak abundances are usually interpreted as
the signatures of low star formation rates, or of an ISM enriched
by an early generation of stars that is now evolving quiescently.
Therefore, the abundances observed here are reminiscent of the 
truncated activity observed in lower $z$ clusters.

At intermediate redshifts ($z \sim 0.5$),
observations indicate that  star formation is suppressed in rich 
cluster galaxies (possibly after an initial burst of enhanced activity) 
by processes such as ram pressure and tidal stripping (Couch et al. 2001 and
references therein).  Just as this process manifests itself as
an excess of post-starburst (E + A) galaxies with no strong emission
lines but strong Balmer absorption, so we may expect to see the
chemical signature of star formation truncation.  However, it
would be somewhat surprising if such processes already have such
an effect on star formation at $z \sim 2$ where environments
are relatively poor and canonical rich clusters have yet to
form.  In addition, although the kinematics of these DLAs, determined
from the unsaturated metal lines, extend over $\sim$ 100 \kms,
this is not atypical of the range exhibited by other DLAs, i.e.
there is no evidence for significant disturbance of the ISM.

Follow-up observations of this field, by either multi-colour (to
determine photometric redshifts) or narrow band \lya\ imaging are
clearly of great interest to confirm the richness of the environment
around these DLAs.  In addition, it is important
to identify other multiple DLAs and follow them up with high
resolution spectroscopy in order to ascertain whether DLAs in
groups exhibit distinct abundances from their `field' counterparts.

\begin{acknowledgements}

S. L. acknowledges financial support by FONDECYT
grant N$^{\rm o} 3\,000\,001$ and by the Deutsche Zentralstelle f\"ur
Arbeitsvermittlung.  We are grateful to Marcin Sawicki for useful 
discussions and to Max Pettini and Jason Prochaska (the referee) for
comments and suggestions that have helped to improve this work.

\end{acknowledgements}

\end{document}